\documentclass[journal=jpcbfk,manuscript=draft,layout=traditional]{achemso}

\usepackage{chemformula} % Formula subscripts using \ch{}
\usepackage[font=small,labelfont=normalfont,singlelinecheck=false,justification=raggedright]{caption}
\usepackage{blindtext}
\newcommand{\Savg}{S$_{avg}$ }
\newcommand{\Smin}{S$_{min}$ }
\newcommand{\Spl}{S$_{PL}$ }

\author{Fabian Keller}
\author{Andreas Heuer}
\email{andheuer@uni-muenster.de}
\phone{+49 251 83-29177}
\affiliation[Westfälische Wilhelms-Universität]{Institute of Physical Chemistry, University of Münster, Corrensstraße 28, 48149 Münster, Germany}

\title{A mechanistic perspective on the effect of cholesterol in phospholipid bilayers}

\begin{document}

%%%%%%%%%%%%%%%%%%%%%%%%%%%%%%%%%%%%%%%%%%%%%%%%%%%%%%%%%%%%%%%%%%%%%
%% The "tocentry" environment can be used to create an entry for the
%% graphical table of contents. It is given here as some journals
%% require that it is printed as part of the abstract page. It will
%% be automatically moved as appropriate.
%%%%%%%%%%%%%%%%%%%%%%%%%%%%%%%%%%%%%%%%%%%%%%%%%%%%%%%%%%%%%%%%%%%%%
%\begin{tocentry}
%\includegraphics[]{./figures/toc_img2.png}
%\end{tocentry}

%%%%%%%%%%%%%%%%%%%%%%%%%%%%%%%%%%%%%%%%%%%%%%%%%%%%%%%%%%%%%%%%%%%%%
%% The abstract environment will automatically gobble the contents
%% if an abstract is not used by the target journal.
%%%%%%%%%%%%%%%%%%%%%%%%%%%%%%%%%%%%%%%%%%%%%%%%%%%%%%%%%%%%%%%%%%%%%
\begin{abstract}
Cholesterol (CHOL) is one of the most important components of plasma membranes of higher cells and one of the main factors for the formation of (nano)domains. In this work,  molecular dynamics simulations of mixtures of CHOL with DPPC (saturated lipid) and DLiPC (unsaturated lipid) as standard phospholipids (PLs) are presented in a wide range of CHOL concentrations. The key idea is to systematically extract all structural and enthalpic properties relevant to the formulation of a lattice model of these systems and express them in dependence of the acyl chain order parameters. Detailed interpretation is simplified by the observation that, to a good approximation, the interaction effects do not depend on the total CHOL concentration, but only on the local CHOL arrangement. The resulting information can be used to motivate the agglomeration of CHOL molecules, the relevance of entropic rather than enthalpic effects for understanding the stronger influence of CHOL on DPPC compared to DLiPC, or the thermodynamic background of raft formation. It is verified that the interaction functions hardly change during the transition from binary to ternary mixtures, suggesting the applicability of the concepts to more complex mixtures.
\end{abstract}

%%%%%%%%%%%%%%%%%%%%%%%%%%%%%%%%%%%%%%%%%%%%%%%%%%%%%%%%%%%%%%%%%%%%%
%% Start the main part of the manuscript here.
%%%%%%%%%%%%%%%%%%%%%%%%%%%%%%%%%%%%%%%%%%%%%%%%%%%%%%%%%%%%%%%%%%%%%
\section{Introduction}
The prominent role of CHOL in the formation of functional nanodomains has been widely accepted.\cite{Levental2020, Sezgin2017, Jacobson2007, Hancock2006, Edidin2003} Since its initial formulation as lipid rafts\cite{Simons1997}, many efforts have been made to gain a mechanistic understanding of their properties in live cells and their associated biological function. The nanodomains are associated with several vital membrane functions, like trafficking and signaling, activation, and quenching of proteins (through enrichment), and are proposed to be an attack site for pathogens.\cite{Simons2011, Rajendran2005, VanDerGoot2001} Surprisingly, nanodomains could, to date, not be directly observed in cell membranes due to resolution limits in time and length.\cite{Sigal2018, Hancock2006, Pike2006, Honerkamp-Smith2008, Levental2016} Computational models are, therefore, of great value for the understanding of the underlying mechanisms of nanodomain formation. The problem of molecular dynamics simulations is the limitation in the compositional complexity of live cells, even though some breakthroughs have been made.\cite{Ingolfsson2014, Friedman2018} Additionally, the nonequilibrium state that characterizes life poses a problem for the formulation of model systems.\cite{Fan2010} Simplified (lattice) models reduce a complex relation into simpler sub-issues, each reproducing a specific property of the complex relationship. Thus, each model adds a facet to the big picture. Concepts to be elucidated included the determination of coexisting domains as a function of composition and temperature, domain sizes and their fluctuations, the understanding of the properties and the resulting impact of CHOL in membranes, or the lateral isotropy of lipid interactions (e.g., for hybrid lipids, or CHOL's smooth and rough face).

Lattice models can be characterized by their design choices for the lattice geometry, the interaction parameters, and the representation of each site.
Neutron scattering experiments\cite{Armstrong2013, Toppozini2014} and atomistic MD simulations\cite{Sodt2014, Sodt2015a, Zhang2016, Javanainen2017} have shown that acyl chains form hexagonal patterns which might guide the choice of the lattice geometry. The hexagonal/triangular geometry, indeed, was utilized in many models that were formulated to reproduce properties of CHOL-like phase behavior\cite{Almeida2009, Tumaneng2011, Dai2011, Ehrig2011}, nano domain size\cite{Frazier2007, Gomez2008, Yethiraj2007}, and segregation behavior\cite{Wang2018, Pandit2007, Shlomovitz2014, Palmieri2013a}. Most of these lattice models use pairwise interactions. The parameters for the interaction functions are derived from experiments or tuned to give correct system behavior.
To the best of our knowledge, all these models require  phenomenological input to properly fit CHOL characteristics and typically use a constant PL-CHOL interaction.

In previous work, our group has developed a lattice model for mixtures of the phospholipids (PL) DPPC (saturated PL) and DLiPC (unsaturated PL) \cite{Hakobyan2017a,Hakobyan2019}. One key aspect was the consideration of the PL-chain order parameter $S$  as one degree of freedom. The model development occurred in two steps. At level I,  the {\it neighbor functions}, expressing the number of neighbors as a function of $S$ and the {\it interaction functions} (equivalently: {\it enthalpy} function), reflecting the pairwise lipid-lipid interactions, again in dependence of the chain order parameters,  were extracted from molecular dynamics (MD) simulations of the DPPC systems, the DLiPC systems, and the DPPC-DLiPC mixtures. Note that these energies incorporate all structural effects that describe lipid ordering (e.g. spontaneous curvature, which was reported to be a significant factor for lateral structuring asymmetric bilayers\cite{Allender2020}), and represent the mean properties of such systems. This information alone provided a new perspective on the properties of these systems and a qualitative understanding of the different aspects of the system. At level II, the $S$-dependent chain entropy was adjusted in an iterative procedure in such a way that the Monte Carlo simulations of the lattice model finally yielded the same distribution of order parameters as in the MD simulations. With this complete model different aspects could be positively compared with the results of MD simulations, such as the reproduction of the phase transition temperature or the time dependence of the phase separation dynamics in case of DPPC-DLiPC mixtures. Furthermore, a detailed comparison between atomistic and coarse-grained force fields was possible,  revealing the fine-tuned interplay of enthalpy and entropy \cite{Hakobyan2019}.

In recent work it has been shown that the properties of PL-CHOL mixtures are mainly driven by the the nearest neighbor PL-PL and PL-CHOL interactions as well as entropic effects \cite{Keller2021}. In contrast, interleaflet and water interactions hardly varied with the CHOL concentration, indicating that interleaflet do not play a role in the ordering effect of CHOL. Based on this insight, in this work we formulate a lattice model for PL-CHOL mixtures which, on the one hand, conceptually is similar to our previous work on pure PL-systems but, on the other hand, due to the major structural differences of CHOL and PLs, involve major modifications as compared to the pure PL-based system. CHOL concentrations up to 30\% are analyzed.  Here we restrict ourselves to the level I which already allows a closer understanding of the specific impact of CHOL on DPPC or DLiPC. It provides answers to, e.g., the following questions: To which degree can the well-known condensing effect of CHOL be taken into account? Is the PL-PL interaction significantly modified by the presence of CHOL? Is it possible to understand the ordering effect of PL due to the presence of CHOL? Is it possible to characterize the system by neighbor and interaction functions which can be used for all CHOL concentrations? The last question is of particular relevance because a positive answer would strongly increase the applicability of a lattice model, e.g., to study domain formation on the mesoscale for which locally averaged CHOL concentrations may vary quite significantly within the overall system.

\section{Computational details}

\subsection{MD simulations}

All bilayers containing random mixtures of DPPC, DLiPC, and CHOL were prepared using the CHARMM-GUI web-based graphical interface.\cite{Lee2016}
The simulations of the DPPC bilayers, the DLiPC bilayers, and the PL/CHOL bilayers were run for 1~$\mu$s. The initial 50~ns of each trajectory were discarded to assure the exclusive analysis of equilibrium data. The simulations were run at temperatures from 290~K to 350~K in 10~K steps. Bilayer compositions and trajectory lengths are listed in table \ref{tab:MDlist} in the SI. 

All simulations were conducted using the CHARMM36 force-field\cite{Klauda2010,Lim2012a} and Gromacs~2020 software package\cite{Pall2015,Abraham2015}. The systems were equilibrated using the 6-step established CHARMM-GUI parameter set, gradually decreasing restraints on lipid head group positions and chain dihedral angles (see table \ref{tab:equilibration_protocol}). The TIP3P model was used as water model.\cite{Jorgensen1983} The Nos\'e-Hoover algorithm was used to maintain the temperature with a coupling constant of 1~ps, coupling bilayer and solvent separately.\cite{Nose1984} To maintain the pressure at 1~bar the Parrinello-Rahman barostat was used with a semiisotropric coupling scheme, a coupling constant of 5~ps and compressibility of 4.5x10$^{-5}$ bar$^{-1}$.\cite{Parrinello1981} Particle mesh Ewald electrostatics were used with a real-space cutoff of 1.2~nm.\cite{Darden1993} The Lennard-Jones potential was shifted to zero between 1.0 and 1.2~nm, with a cutoff of 1.2~nm and the nonbonded interaction neighbor list was updated every 20 steps with a cutoff of 1.2~nm.

\subsection{Parameter definitions and software}

Handling and analysis of MD data heavily relied on the MDAnalysis package for Python \cite{Michaud-Agrawal2011b,Gowers2016}. 
All molecular visualizations in this work were created using VMD \cite{Humphrey1996}.

The radial distribution functions (RDFs) were calculated using the Gromacs analysis tools \cite{Abraham2015}. All RDFs were calculated individually for each leaflet using in-plane distances and if not stated otherwise were averaged. For the PL molecules, the phosphor atoms and, for CHOL, the hydroxyl oxygen atoms were used as reference positions.

The lipid chain order parameter S is defined as $S = \left< 1.5\cos^2\theta - 0.5 \right> $, where $\theta$ denotes the angle between the vector spanned by every second carbon atom in a lipid chain and a the bilayer normal (z-axis). This definition of calculating the order parameter differs from the commonly used S$_\mathrm{CH}$ order parameter, as we do not consider hydrogen atoms in this calculation. Thus $S=-0.5$ is the lowest order state, while $S=1.0$ is the highest order state (linear chains). This has the advantage of easier comparability between the parameters of saturated and unsaturated PLs.

\section{Results and discussion}

\subsection{Lateral bilayer geometry of binary PL/CHOL mixtures}

 %%%%%%%%%%% %%%%%%%%%%% %%%%%%%%%%% %%%%%%%%%%%
%%%%%%%%%%%%% RDFs
 %%%%%%%%%%% %%%%%%%%%%% %%%%%%%%%%% %%%%%%%%%%%
 
\begin{figure}[bt]
    \centering
    \includegraphics[width=0.9\textwidth]{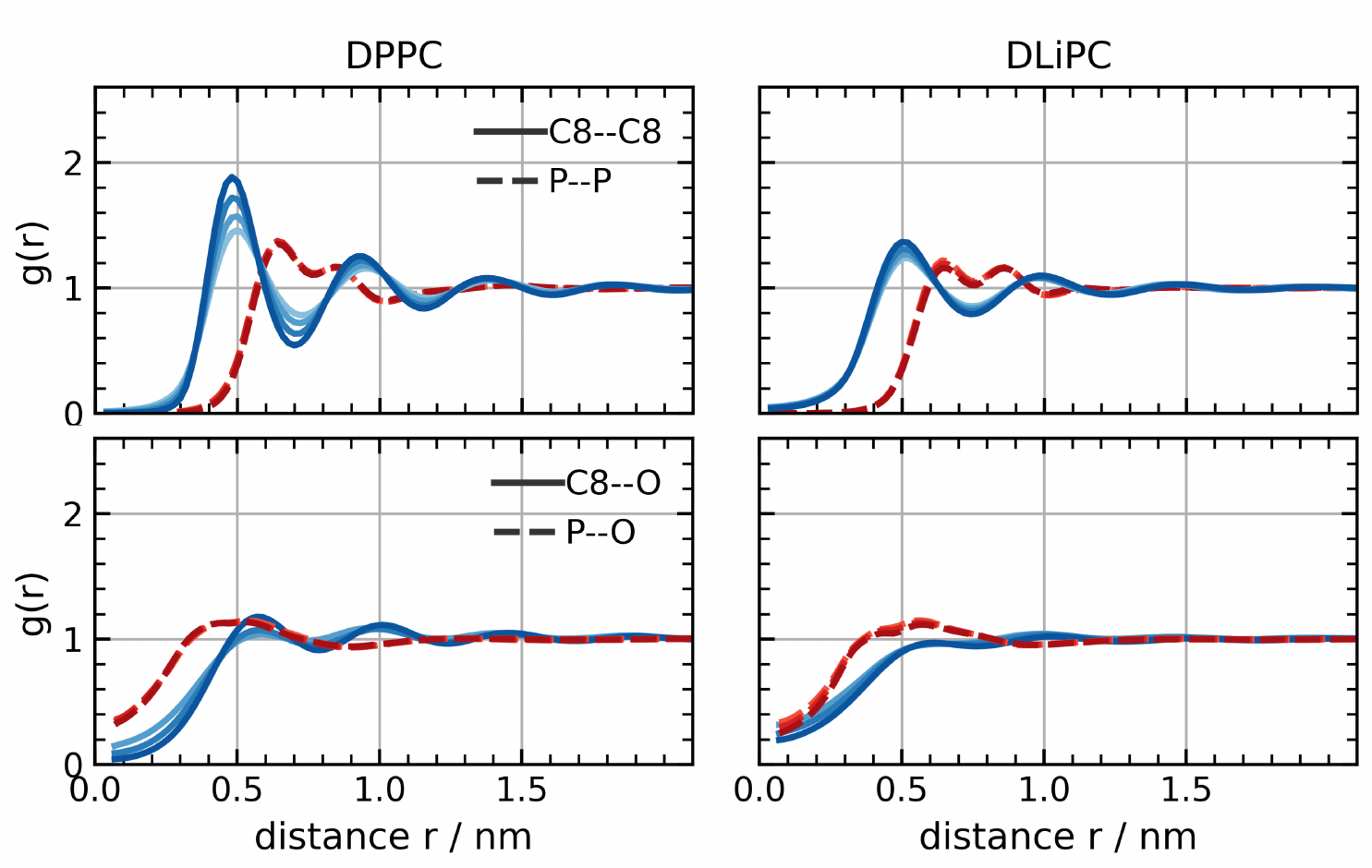}
    \caption{Radial distribution functions of PL-PL (top) and PL-CHOL (bottom) lateral distances from simulations of DPPC/CHOL or DLiPC/CHOL (with CHOL concentrations from 0 to 30\%) averaged over simulations at temperatures above the phase transition. For the PL head groups phosphor atoms (dashed lines), for the CHOL head groups the hydroxy oxygen atoms and for the PL tails the eighth carbon atoms (solid lines) were used as reference positions. Shades of blues (tail level) or reds (head group level) indicate the CHOL concentration, with darker colors representing higher CHOL concentrations. }
    \label{fig:rdfs_pl}
\end{figure}

First, we analyse radial distribution functions (RDFs) which describe the local structure and are the basis to setting up the geometry of the lattice. Here, we show in figure \ref{fig:rdfs_pl} the RDFs of PL-PL and PL-CHOL lateral distances at the head group level (P-P and P-O atom distances) and at the acyl chain level (C8-C8 and C8-O atom distances) averaged over the respective simulations above the phase transition temperatures. The RDFs are used for a consistent definition of nearest neighbors. Note that the PL-CHOL RDFs are finite at r=0 because the hydroxy oxygens of CHOL lie in a different bilayer height than the C8 or P atoms and can thus overlap.

The RDFs of PL-PL distances show profound differences comparing head group and chain level. The chains are well-structured, up to distances of 1.5~nm (second neighbor shell, see \cite{Leeb2018a}). The structuring is stronger with increasing CHOL concentration, especially for DPPC. In contrast, the CHOL concentration exhibits no influence on the head group structure (figure \ref{fig:rdfs_pl} top) and accordingly we find that the chain level is more sensitive to changes of the bilayer composition.

The DPPC-CHOL RDFs (figure \ref{fig:rdfs_pl} bottom) show very similar behavior compared to the head and tail levels of the PL-PL. Generally, the structuring between DPPC and CHOL is weaker than between DPPC and DPPC. The chain level RDFs also show increased structuring by CHOL, while CHOL does not affect the head level structure. Interestingly, the maxima of the DPPC-CHOL chain level RDFs roughly coincide with the DPPC chain level maxima, indicating that CHOL takes the position of an acyl chain. The RDFs of DLiPC-CHOL, on the other hand, show no structuring in the chain region.

The investigation of the different structuring behavior of the head and tail levels through the RDFs showed that, contrary to the chain region, the head group structures are independent of the concentration of CHOL. Accordingly, a consistent neighbor definition using a static distance cutoff is possible employing the head groups as reference positions for, e.g., lattice models. Based on the first minimum of PL-PL and PL-CHOL RDFs we use a distance cutoff of 1~nm between the head group atoms to define nearest neighbors in this work.

 %%%%%%%%%%% %%%%%%%%%%% %%%%%%%%%%% %%%%%%%%%%%
%%%%%%%%%%%%% NPLPL %%%%%%%%%%%
 %%%%%%%%%%% %%%%%%%%%%% %%%%%%%%%%% %%%%%%%%%%%
 
\begin{figure}[bt]
    \centering
    \includegraphics[width=0.9\textwidth]{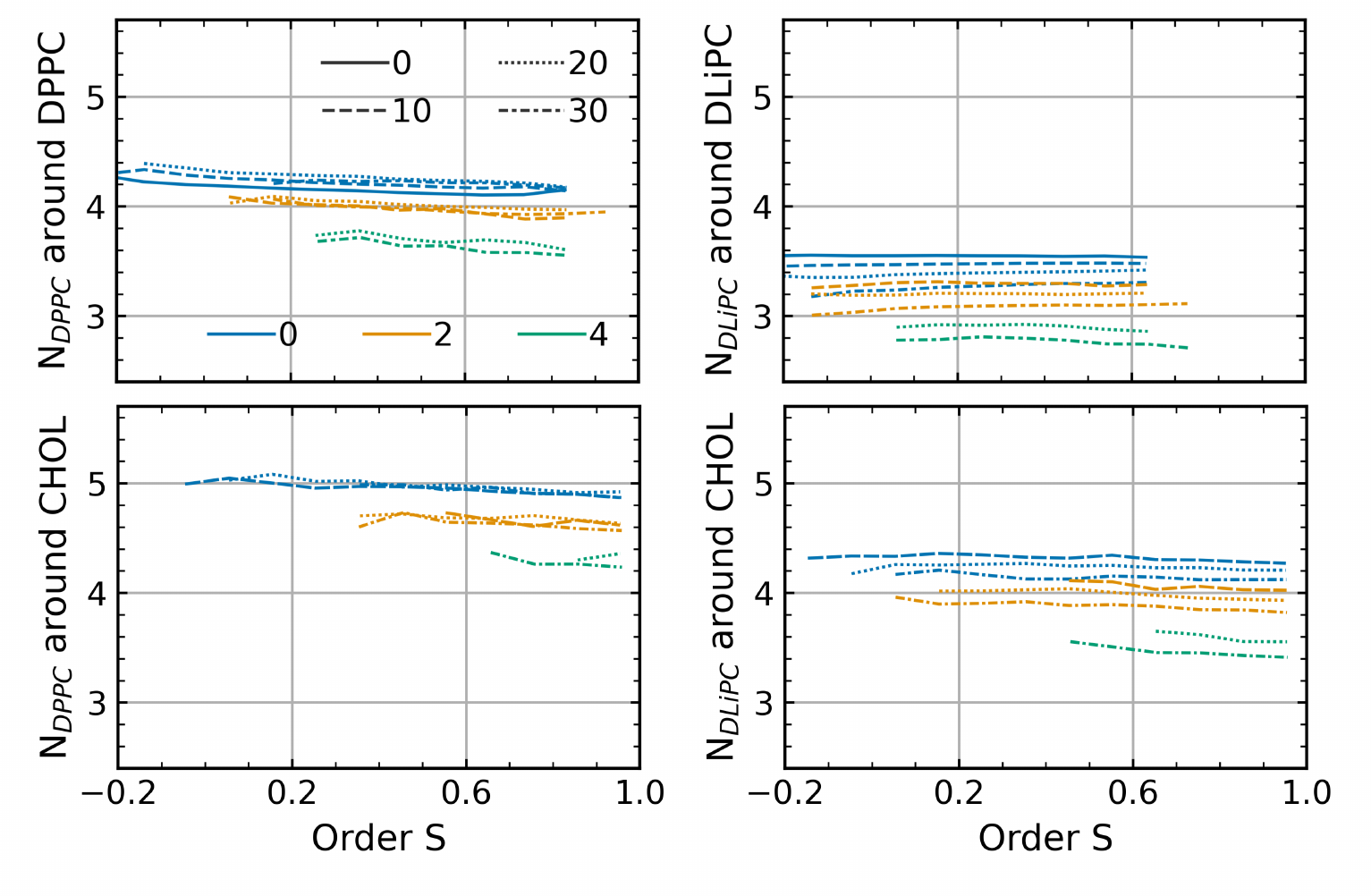}
    \caption{Top: Number of nearest DPPC (left) and DLiPC (right) neighbors of a PL as a function of its order parameter, separately evaluating reference lipids with 0, 2, and 4 nearest CHOL neighbors. Bottom: The respective number of nearest DPPC (left) and DLiPC (right) neighbors around CHOL. The line style indicates the CHOL concentration of the respective PL/CHOL bilayers.}
    \label{fig:NofS}
\end{figure}

Now we can quantify the number of nearest neighbors of PLs or CHOL. The average number of PL neighbors of the respective PL with 0, 2, or 4 CHOL neighbors as a function of its order parameter is shown in figure~\ref{fig:NofS} top. The PL neighborhood around CHOL is shown in figure \ref{fig:NofS} bottom. Note that in our definition, the number of nearest neighbors only reflects the adjacent DPPC and DLiPC lipids, respectively, but {not} CHOL. 

The number of neighbors is not affected by the lipid's order. We find roughly 4.5 and 3.5 neighbors for DPPC and DLiPC, respectively. Interestingly, the number of nearest neighbors hardly depends on the number of CHOL molecules around DPPC; the average number of PL neighbors is decreased by, on average, 0.5 PL neighbors comparing PLs with 0 and 4 CHOL neighbors, regardless of the overall CHOL concentration. From this observation, we can obtain three conclusions: (1) The influence of CHOL on the lateral PL head group structure is weak, reflecting the well-known condensing effect. (2) The residual dependence mainly depends on the local concentration which allows the use of the same neighbor function for all CHOL concentrations. (3) The neighbor function hardly depends on the order parameter which further simplifies the impact of the neighbor function. 

The PL neighborhood around CHOL resembles the neighborhood of DPPC and DLiPC, except that CHOL has, on average, one additional PL neighbor. This increase may be related to the fact that there exists a finite probability that the projection of the CHOL hydroxy oxygen atom on the membrane plane may be close to the projection of the PL head-group. However, all three conclusions, drawn for the PL-PL neighborhood, hold here as well.

 %%%%%%%%%%% %%%%%%%%%%% %%%%%%%%%%% %%%%%%%%%%%
%%%%%%%%%%%%% Model %%%%%%%%%%%
 %%%%%%%%%%% %%%%%%%%%%% %%%%%%%%%%% %%%%%%%%%%%

\begin{figure}[bt]
    \centering
    \includegraphics[width=0.25\textwidth]{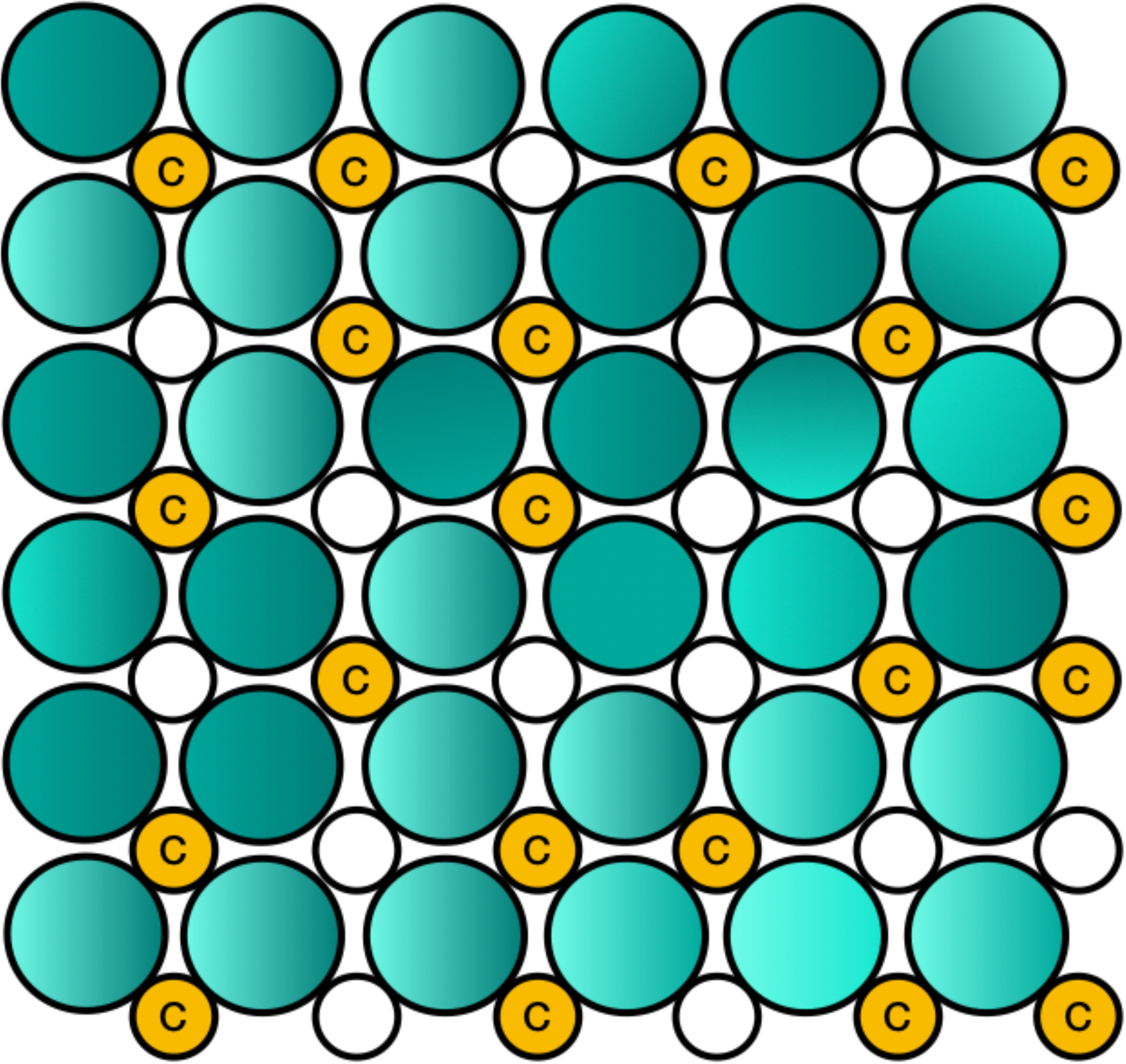}
    \caption{Square lattice representation with CHOL on a sublattice (yellow dots) with void sites. PLs reside on the main lattice (cyan). The shades of cyan indicate the different order parameters that a PL can exhibit.}
    \label{fig:squarelattice}
\end{figure}

Using the presented results, we are in the position to find a lattice geometry, based on the lipid head group region, that can incorporate the average lipid neighborhood. The PL head group region is not affected by the presence of CHOL. Therefore, PLs and CHOL should reside on separate square lattices. In that framework, the CHOL lattice is a sub-lattice with void sites, enabling the simulation of different CHOL concentrations.  A minimum version of the resulting lattice model could be formulated in the way depicted in figure \ref{fig:squarelattice}.
The slight deviation from an optimal number of 4 neighbors on a square lattice can be handled by including an appropriate scaling factor $N/4$ ($N$: total number of neighbors for the specific CHOL environment) for the individual Pl-PL interaction functions, thereby correctly accounting for the total interaction. This minor correction was already successfully used in previous work  \cite{Hakobyan2017} and brings in a very minor mean-field character into the theoretical description.

From the RDF plots of the chain level it is obvious, that the number of PL (chain) neighbors around PL increases with increasing CHOL concentration. Thus, it would no longer be possible to define neighbor functions of chains without taking into account the overall CHOL concentration.

\subsection{PL-PL related interaction functions}

The interaction function between lipids, which form the key aspect of the lattice model,  is influenced by many factors (e.g., variation of lipid head groups, chain length, and saturation). These factors all influence the inherent order of a lipid. In previous work\cite{Keller2021}, we evaluated the interaction of a PL molecule with its environment and expressed the {\it total} interaction as a function of the PL's order parameter $S$. Indeed, a significant dependence on $S$ was observed. Here we go one step further and characterize the interaction between the individual adjacent lipid pairs with indices $i$ and $j$ as a function of the pair's chain order parameters $S_i$ and $S_j$. In particular, we are interested in the effect of CHOL on this interaction which should reflect CHOL's unique ability to alter the phase behavior in lipid mixtures. The interaction functions are determined as averages over the temperature ranges from 330 to 350~K (DPPC) and  290~K to 350~K (DLiPC), respectively.

\begin{figure*}[tbhp]
    \centering
    \includegraphics[width=0.9\textwidth]{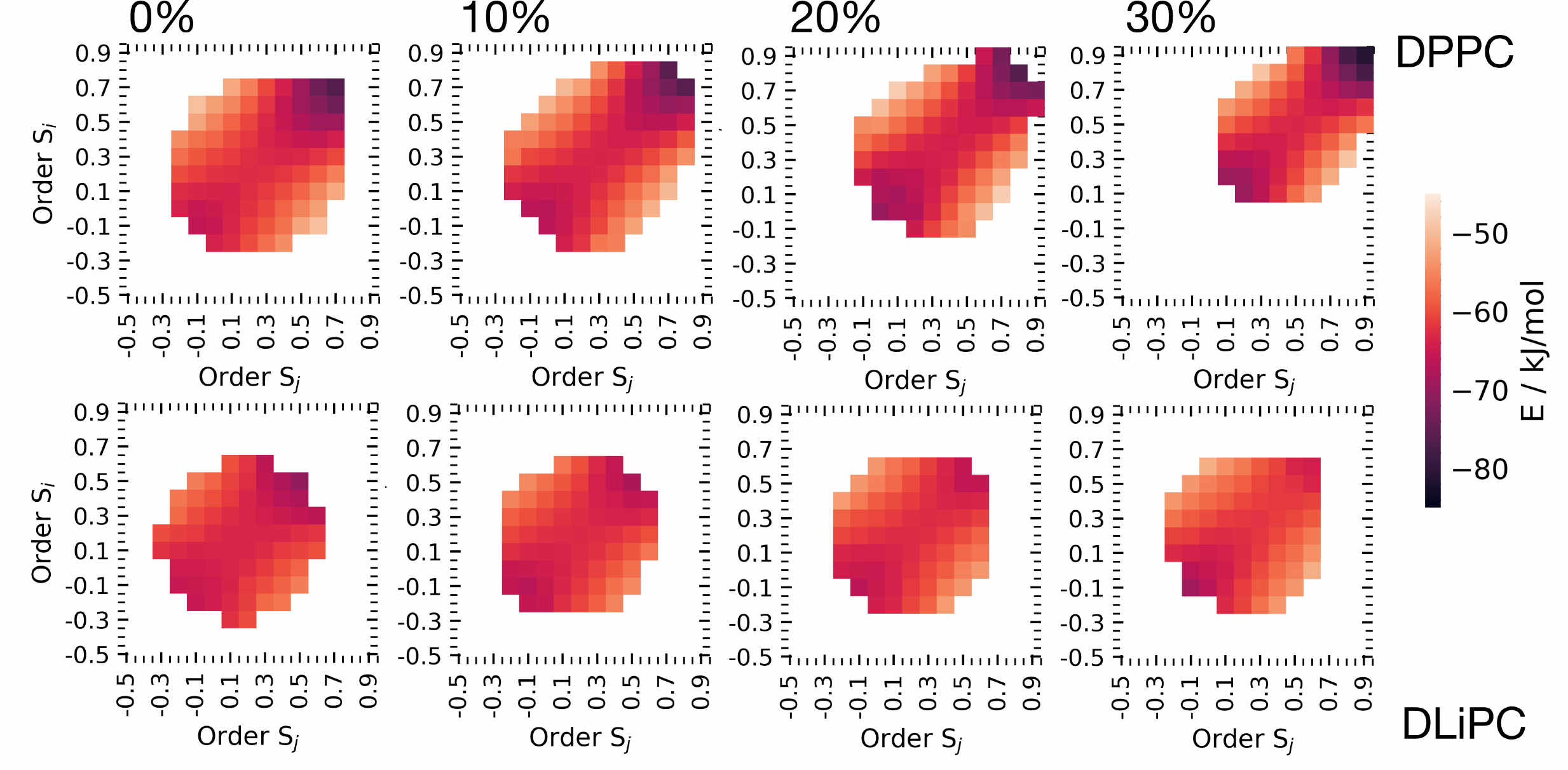}
    \caption{Interaction energy of a pair of lipids i and j of DPPC (top) or DLiPC (bottom) as a function of the order parameters S$_i$ and S$_j$ in bilayers of DPPC/CHOL or DLiPC/CHOL at CHOL concentrations from 0 to 30\%.}
    \label{fig:interaction2d}
\end{figure*}

%%%%%%%% 2D map %%%%
The interaction energy matrix $E(S_i, S_j)$ of a pair $ij$, shown in figure \ref{fig:interaction2d}, is necessarily symmetric. The interaction increases with increasing order parameters, and the CHOL concentration does not change the general features of the energy matrix. Following our previous work, we aim to express the  interaction matrix as a 1D function where the argument depends on both order parameters. This approach is convenient from a statistical perspective (less noisy) and allows a more direct view on the dependence of the interaction energy on the order parameter. Of course, when actually approaching level II (employing \textit{neighbor} and \textit{enthalpy} functions to get the \textit{entropy}) of the lattice modeling, one might still have the option to work with the full 2D interaction matrix. We now have multiple options to simplify the interaction description (i.e., reducing the two-dimensional to a one-dimensional function). In previous work \cite{Hakobyan2017} we chose the pair's average order parameter \Savg for PL mixtures. For this purpose we  calculated the weighted average for given average order parameter. Here we propose another scheme, using the minimum order parameter \Smin averaging over horizontal/vertical entries in the interaction matrix and show that this choice is superior to our previous one.

\begin{figure}[tbhp]
    \centering
    \includegraphics[width=0.9\textwidth]{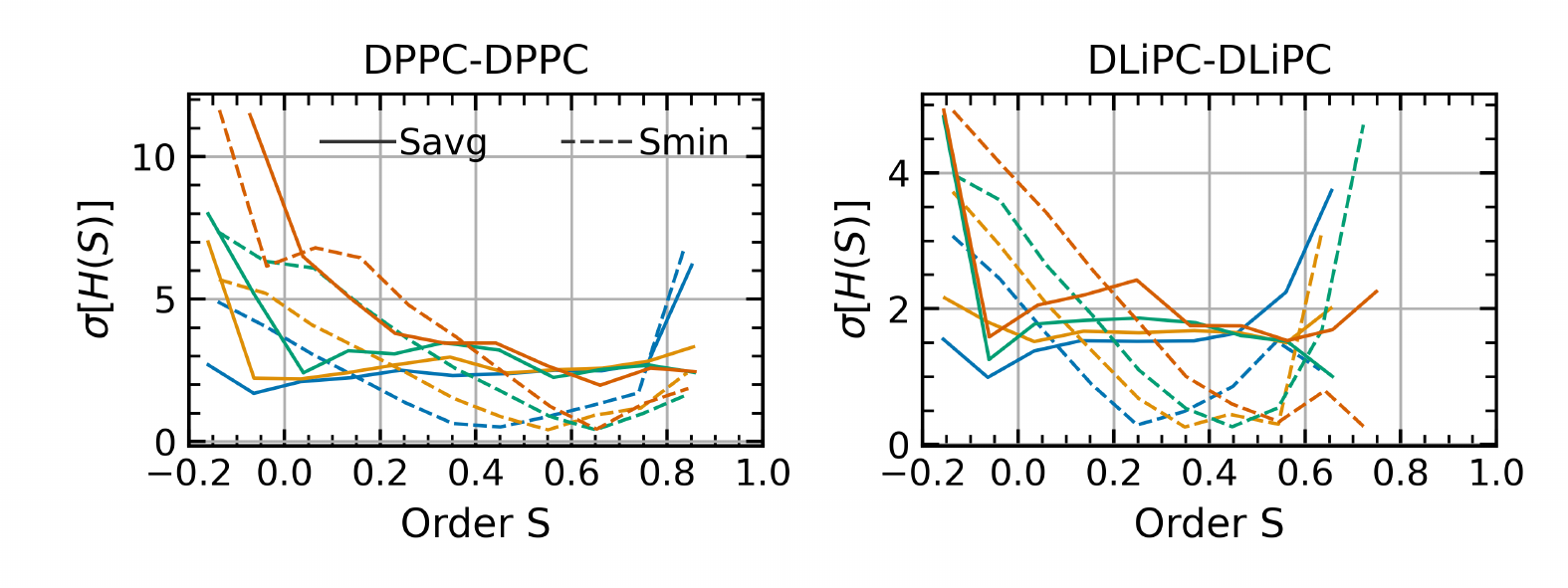}
    \caption{Standard deviation of the pair interaction energies if the minimal order parameter of the lipid pair is considered (Smin, horizontal/vertical averaging of \ref{fig:interaction2d}) or if the average order parameter of the pair is considered (Savg, diagonal averaging). Blue, yellow, green and red indicate CHOL concentrations of 0, 10, 20 and 30\%, respectively.}
      \label{fig:s_scheme_var}
\end{figure}

An optimal choice minimizes the error of using the 1D variant as compared to the complete 2D interaction matrix.  Each CHOL concentration is analyzed separately.  Specifically, we calculated for each value of the respective order parameter (\Savg or \Smin) the weighted variance, see figure \ref{fig:s_scheme_var}, where its square root, i.e. the standard deviation, is displayed. It shows that in the relevant regime of positive order parameters, \Smin is a better choice than \Savg.  Furthermore, when weighting each value with the occurrence probabilities of the individual order parameters, one obtains an overall quality (figure \ref{fig:s_scheme_var_bar}) which also shows the superiority of the \Smin-scheme, in particular for DPPC.

\begin{figure}[htbp]
    \centering
    \includegraphics[width=0.9\textwidth]{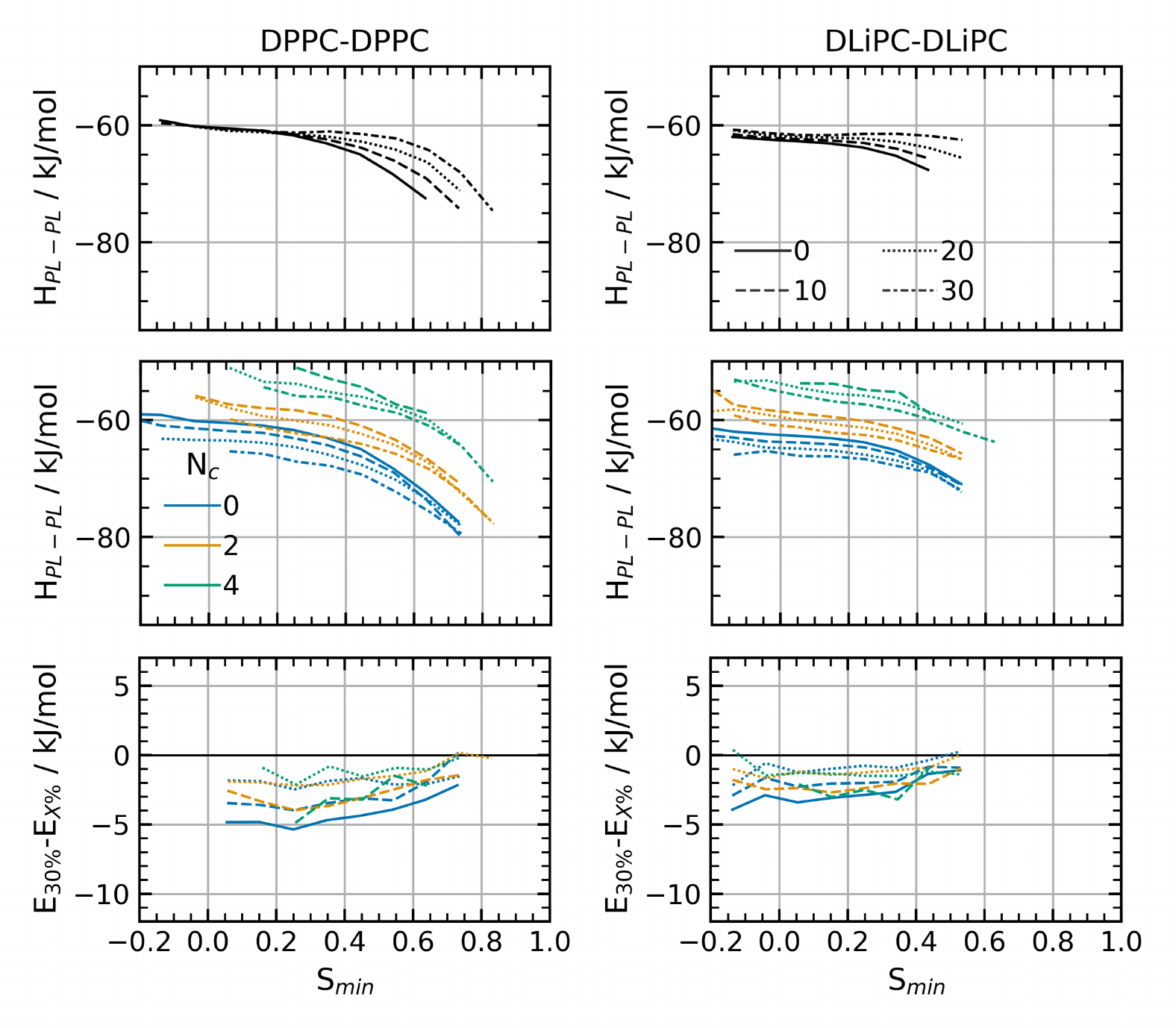}
    \caption{Top: PL pair interaction energies as a function of the pair’s minimal order parameter and derived from binary PL/CHOL mixtures with CHOL concentrations from 0 to 30\%. Middle: The respective PL pair interaction energies as a function of the pair’s minimal order parameter when separated by the number of the pair’s nearest CHOL neighbors. N$_c$ considers the total number of CHOL molecules around the pair. Bottom: PL pair interaction energy difference between energies from simulations with 30\% CHOL and 0\%, 10\%, and 20\%.}
\label{fig:Eplpl}
\end{figure}

In figure \ref{fig:Eplpl} top we show the interaction enthalpy between the PL-PL pairs as a function of \Smin for different CHOL concentrations. In agreement with previous work on DPPC bilayers without CHOL \cite{Hakobyan2017}, the interaction of DPPC molecules is stronger for chains with high order (figure \ref{fig:Eplpl} top).  The interaction energy reaches a plateau at around 60~kJ/mol for \Smin smaller than a threshold value that depends on the CHOL concentration. The DLiPC-DLiPC pair interactions behave quite similarly. Importantly, the PL-PL interaction becomes weaker for higher CHOL concentration. The interpretation is straightforward: since one PL also wants to optimize its interaction with adjacent CHOL molecules (particularly relevant for high CHOL overall concentration), the efficiency of its interaction with the adjacent PL is reduced. 

To get more specific information about the effect of CHOL on the PL-PL interaction, we additionally record the number of direct CHOL neighbors N$_c$ and display the PL-PL interaction in dependence of the overall CHOL concentration {\it and} the value of N$_c$. Interestingly, with this finer resolution the main dependence is with respect to N$_c$. Each additional CHOL neighbor decreases the interaction energy.  In contrast, the explicit dependence on the overall concentration in this representation is quite weak (figure \ref{fig:Eplpl} middle) and displays a slightly stronger interaction strength for higher CHOL concentration. This may reflect the residual global effect of a generally tighter packing in bilayers of high CHOL content. These observations imply that the initially observed dependence on the CHOL concentration (figure \ref{fig:Eplpl} top), reflecting a weighted average of the distribution of N$_c$, is just a consequence of the typically larger number of direct CHOL neighbors for a PL with high chain order. 

These results also imply that the influence of CHOL is mainly local. 
To quantify the non-local contributions, we compare the interaction energies at 30\% CHOL concentration with those at lower concentrations (figure \ref{fig:Eplpl} bottom). Here we see that the dependence on \Smin is very weak (approx. 2 KJ/mol in the relevant regime of order parameters). Energies that are independent of the configuration and thus do not provide driving forces of chain ordering are irrelevant for the membrane lattice model. Thus, the dependence on the overall CHOL content can be neglected. This rationalizes the choice of identical interaction parameters in the whole range of CHOL concentrations. We would like to add that the dependence on the overall CHOL concentration is even smaller for DLiPC. This holds for the absolute energy shift as well as for the dependence on \Smin.

%%% Paragraph on temperature, composition, and PL type dependence of the evaluated interaction functions
The interaction functions were determined as an average from simulations at 330~K, 340~K, and 350~K for DPPC (290 K to 350 K for DLiPC), presupposing the energies H(S) are not temperature-dependent. To evaluate the temperature dependence of respective interactions, we calculated the difference of the interaction functions from 330~K and 350~K. This energy difference for DPPC and DLiPC pair interactions is depicted in figure \ref{fig:deltaEplpl_Tdepscheme} in the SI. The differences between the energies at 330~K and 350~K lie below 2~kJ/mol. Only the interaction energy for the unlikely case of 30\% CHOL and 0~CHOL neighbors shows a higher temperature dependence. Consequently, we infer that the temperature dependence of the interaction as a function of \Smin and N$_c$ is basically temperature independent in the investigated temperature range.

\begin{figure}[tbhp]
    \centering
    \includegraphics[width=0.95\textwidth]{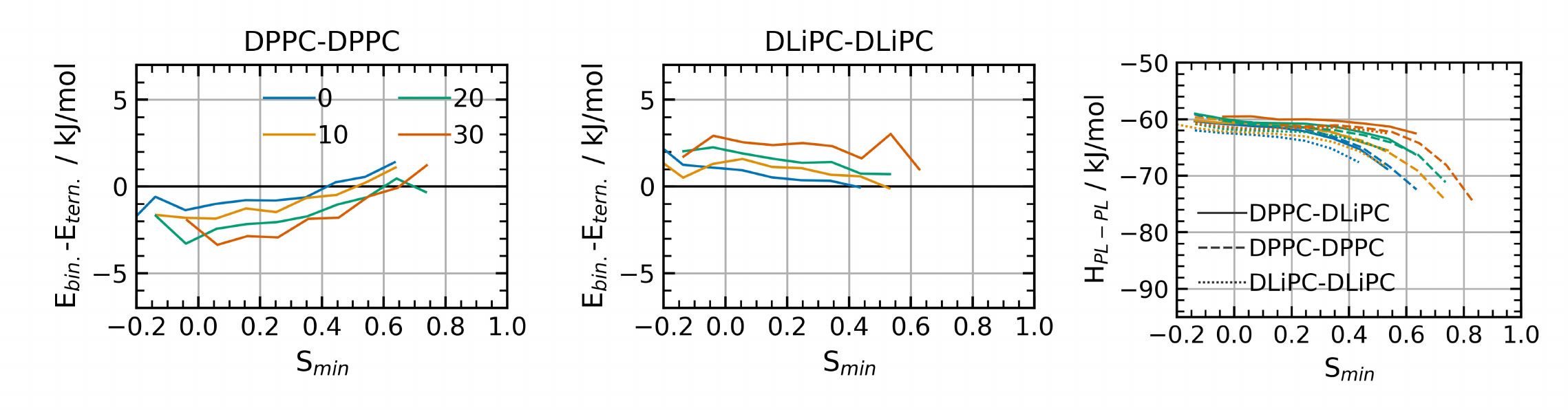}
    \caption{Left and middle: DPPC and DLiPC pair interaction difference between energies derived from binary (PL/CHOL) and ternary (DPPC/DLiPC/CHOL) lipid mixtures as a function of the pair’s minimal order parameter. Right: DPPC-DLiPC pair interaction energies as a function of the pair’s minimal order parameter derived from DPPC/DLiPC/CHOL mixtures with CHOL concentrations from 0 to 30\%.}
    \label{fig:Epl_bintern}
\end{figure}

Finally, we investigate whether the PL-PL interaction profiles depend on the bilayer composition. Therefore, we determined the energy difference between respective PL interactions in binary DPPC/CHOL or DLiPC/CHOL and ternary mixtures of DPPC/DLiPC/CHOL, depicted in figure \ref{fig:Epl_bintern} left. The absolute PL-PL interaction functions are shown in figure \ref{fig:Epl_bintern} right.

For small order parameters the absolute interaction values differ by about 2 kJ/mol between profiles from ternary and binary bilayers whereas they are basically identical in the limit of large order parameters. The interaction in the binary mixtures is stronger for DPPC (decreased packing in ternary mixtures) and weaker for DLiPC (increased packing in ternary mixtures).  Thus, the energy gain for ordering DPPC acyl chains is slightly weakened in the ternary mixtures. However, due to the smallness of these effects we may state the PL-PL interaction is only slightly affected when going from the binary to the ternary bilayer. Thus the present results suggest that the interaction functions, derived from pure PL mixtures, can be, to a good approximation, used also in more complex mixtures.

\subsection{Cholesterol-related interaction functions}

\begin{figure}[tbhp]
    \centering
    \includegraphics[width=0.9\textwidth]{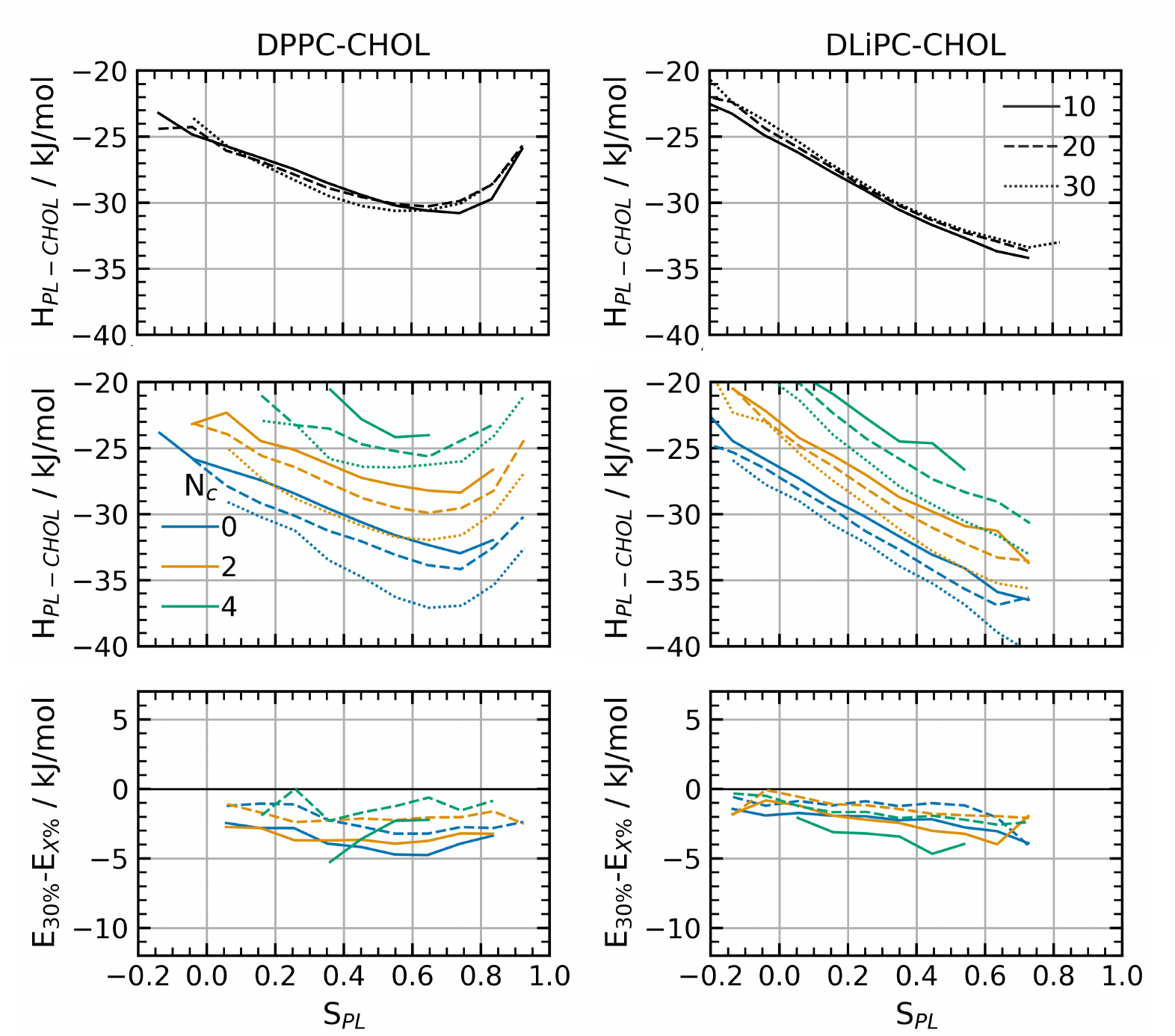}
    \caption{Top: DPPC- and DLiPC-CHOL interaction energies as a function of the PL order parameter S$_{PL}$. The energies were derived from PL/CHOL mixtures with CHOL concentrations ranging from 10 to 30\%. Middle: DPPC- and DLiPC-CHOL interaction energies when separated by the number of the pair’s nearest CHOL neighbors (N$_c$). Bottom: PL pair interaction energy difference between energies from simulations with X=10, 20\% and 30\% CHOL.}
    \label{fig:Eplc}
\end{figure}

%%%%%%%% Introductory %%%%
Next, we investigate the characteristics of the PL-CHOL and CHOL-CHOL interactions. The interaction profile of a PL with CHOL as a function of the PL order \Spl is shown in figure \ref{fig:Eplc} top.  Equivalent to the discussion of the PL-PL interaction functions, we also calculated the interaction for a given  number $N_c$ of CHOL neighbors of the respective PL-CHOL pair (figure \ref{fig:Eplc}  middle) and calculated the differences between the interaction energies from different CHOL concentrations (figure \ref{fig:Eplc} bottom).

Similarly to the PL-PL interactions, the strength of the PL-CHOL pair interaction decreases with each additional CHOL neighbor (figure \ref{fig:Eplc} middle) and, for fixed value of $N_c$ the interaction becomes stronger with increasing CHOL interaction. Both effects are counterbalanced in figure \ref{fig:Eplc} top.

For DPPC and for all CHOL concentrations and $N_c$-values  the strength of the DPPC-CHOL interaction energy displays a maximum for \Spl$\approx$ 0.65. Thus, the optimal PL chain order for interaction with CHOL is not an all-trans but a slightly disordered chain. The DLiPC-CHOL interaction displays a monotonous dependence on \Spl. However, this may be related to the fact that for the unsaturated DLiPC-lipid no order parameters beyond 0.7 are accessible. Interestingly, DLiPC shows a stronger dependence on the order parameter than DPPC which results in a stronger interaction by approx.  4~kJ/mol at \Spl=0.6.  The differences between profiles of different CHOL concentrations lie between 0 and -5~kJ/mol and, importantly, hardly depend on \Spl (figure \ref{fig:Eplc} bottom). In analogy to the PL-PL interaction, the variation with order parameter is less than 2kJ/mol.

\begin{figure}[tbhp]
    \centering
    \includegraphics[width=0.4\textwidth]{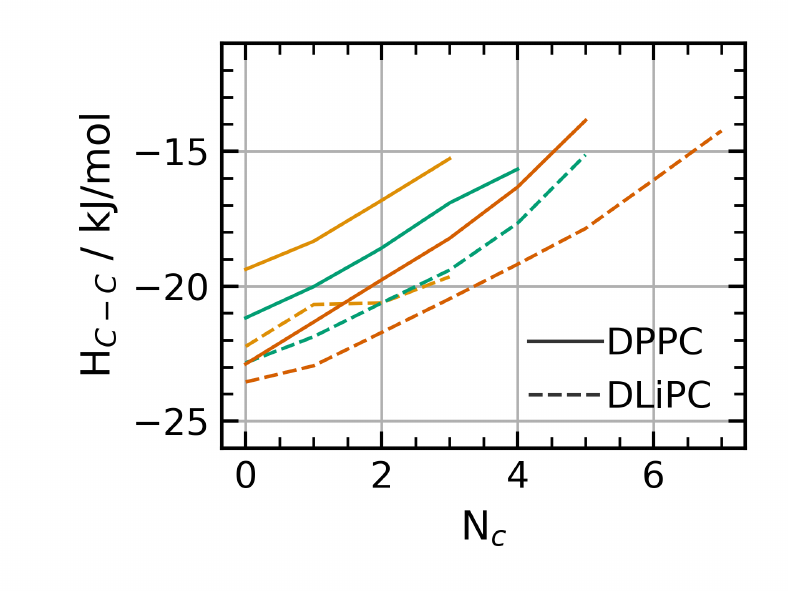}
    \caption{CHOL-CHOL pair interaction energy as a function of the number of their CHOL neighbors N$_c$ derived from simulations of DPPC/CHOL (solid) and DLiPC/CHOL (dashed) mixtures. The CHOL concentration is 10\% (orange), 20\% (green), and 30\% (red).}
    \label{fig:Ecc_Nc}
\end{figure}

%%%%% On the CHOL-CHOL interaction
The CHOL-CHOL interaction does not directly affect the PL order parameter, but it determines how favorable an agglomeration of CHOL is from an energetic point of view. We show the CHOL pair interaction as a function of the pair's CHOL neighbors N$_c$ in DPPC and DLiPC bilayers and the overall CHOL concentration in figure \ref{fig:Ecc_Nc}. Qualitatively, we see the same tendency as for the other two interactions: stronger interaction for less CHOL neighbor and, as a much smaller effect, higher CHOL content. In DPPC mixtures, the concentration dependence is slightly higher with approximately 2~kJ/mol per 10\% CHOL. In DLiPC mixtures, the concentration dependence is roughly half as strong as in DPPC. The interaction decreases with 1 to 1.5~kJ/mol per N$_c$, both in the DPPC and DLiPC mixtures, and the overall interaction strength between CHOL is slightly higher in DLiPC mixtures.

\section{Comparison with MD results}

\subsection{Single PL-CHOL mixture: from the lattice model to MD results }

Here we discuss for interesting properties of CHOL-containing membranes, as obtained from MD simulations, how the results can be interpreted from the properties of the neighbor and interaction functions.

\begin{figure}[tbhp]
    \centering
    \includegraphics[width=0.5\textwidth]{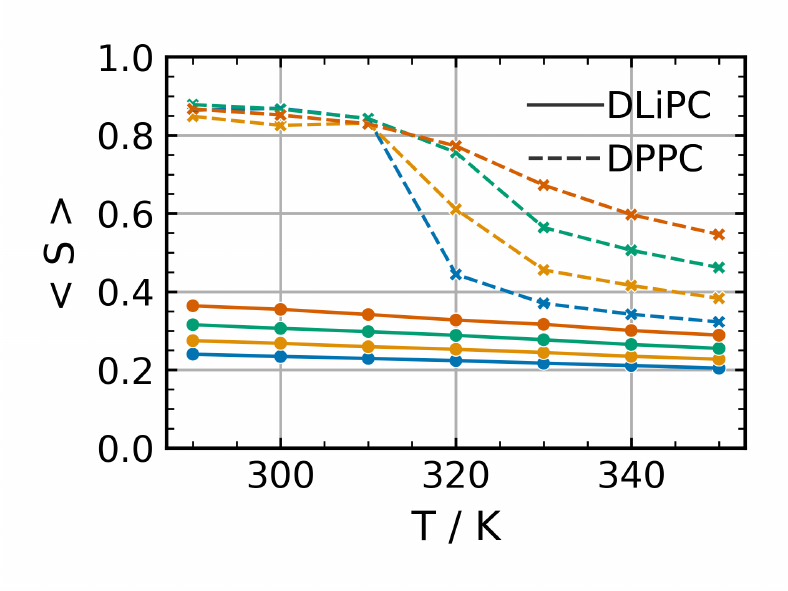}
    \caption{Average order parameter as a function of temperature and total CHOL concentration for DPPC/CHOL and DLiPC/CHOL mixtures. The CHOL concentrations are 0\%, 10\%, 20\%, 30\% (from below to above).}
    \label{fig:averageS}
\end{figure}

We start with the evaluation of the order parameters in the different CHOL-containing bilayers, shown in figure \ref{fig:averageS}. The average order of the DLiPC-containing bilayers increases linearly with CHOL concentration and decreasing temperature. The CHOL ordering is roughly twice as strong for DPPC as it is for DLiPC. For DPPC, we see a phase transition from gel to the liquid phase at around 320~K. The properties of the PL-CHOL interaction function suggest a possible explanation. In the relevant range of order parameters below  \Spl $\approx 0.7$ the PL-CHOL interaction strength increases with increasing order parameter \Spl. Thus, addition of CHOL and the concomitant increasing relevance of the PL-CHOL interaction for the overall energy balance prefers higher PL order parameters.  Since the interaction functions are only determined for temperatures in the liquid phase, we restrict our discussion to this regime. 

\begin{figure}[tbhp]
    \centering
    \includegraphics[width=0.9\textwidth]{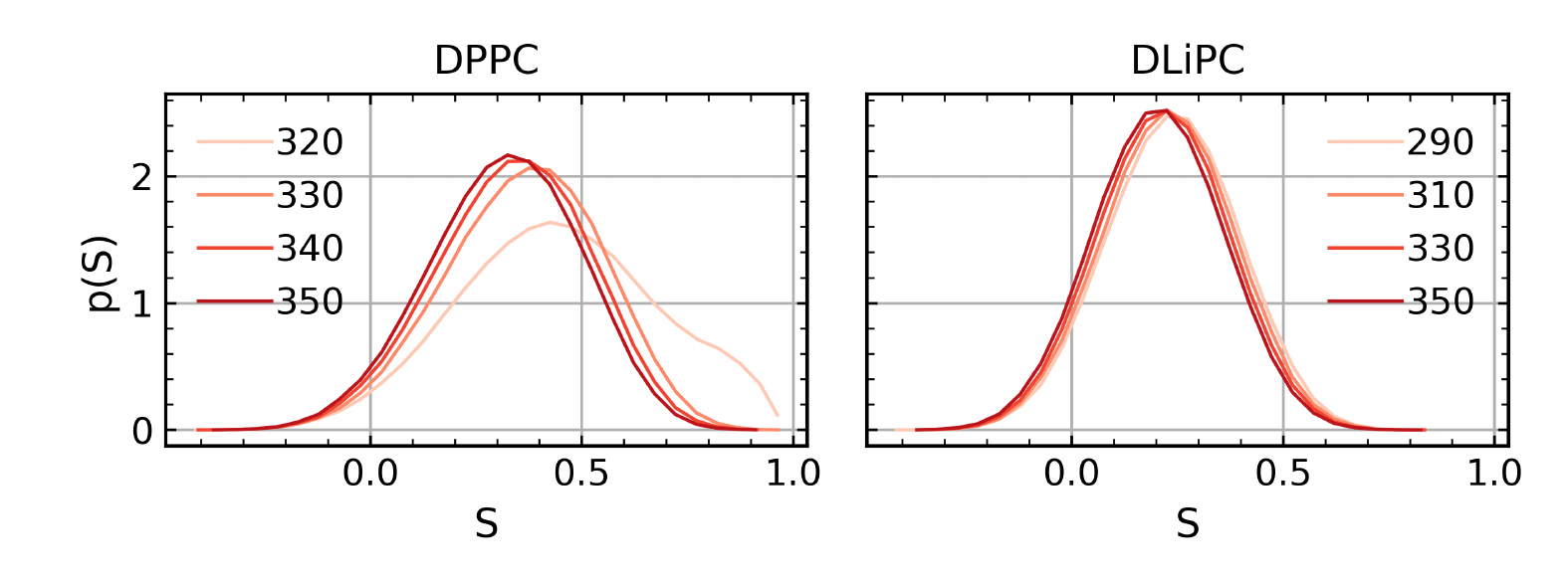}
    \caption{Order parameter distributions of simulations of pure DPPC or DLiPC bilayers close to the phase transition temperature of DPPC ($\approx 320$~K). }
    \label{fig:pscd}
\end{figure}

However, a closer analysis of the MD results reveals several important features, for which a fresh perspective is provided based on the results of this work. (1) Why is the order parameter for the CHOL-free case higher for DPPC ($\approx 0.35$) than for DLiPC ($\approx 0.2$)? The PL-PL interaction functions suggest that the enthalpic gain upon ordering is not stronger than for DLiPC (compare, e.g., the increase of energy between order parameters of 0 and 0.4 in figure \ref{fig:Eplpl} which is approx. a factor of 1.5 larger for DLiPC). Although this effect is slightly counterbalanced by the fact that the  number of nearest PL neighbors is smaller by a factor of $\approx0.75$ for DLiPC bilayers compared to the DPPC bilayers, these considerations do not explain the significant difference between the order parameters of DPPC vs. DLiPC. Thus, this observation suggests that the difference in chain entropy must be used to explain this difference.  Indeed, due to the disordered nature of DLiPC one may envisage that for DLiPC there is a significant entropy penalty to reach more ordered states. Among others, this also reflected by the observation that for DLiPC  order parameters \Smin above 0.5 are hardly visited whereas for DPPC order parameters up to 0.8 can be observed. (2) Why is the impact of CHOL on the chain order much larger for DPPC than for DLiPC? Again, this cannot be explained by the specific form of the PL-CHOL interaction function since, if at all, it would predict a stronger impact of CHOL on the DLiPC order. However, the chain entropy argument, already used in (1), would again explain the additional suppression of states with higher order parameters for DLiPC. In contrast, the enthalpic gain at higher order parameters upon addition of CHOL is less suppressed by the chain entropy, which naturally favors smaller order parameters. (3) Why is the impact of CHOL on the DPPC order particularly strong at $T =320$~K?   We envision that this effect is related to the critical effects close to the transition to the gel state. As shown in figure \ref{fig:pscd} the fluctuations of the order parameter for pure DPPC is particularly large at $T =320$~K. In analogy to the general system behavior at critical points (e.g. for the Ising model close to $T_c$), the system is particularly susceptible to external perturbations such as the addition of CHOL. 

Finally, one may wonder whether this insight may help to rationalize the emergence of rafts for the ternary mixtures, i.e. the presence of a DLiPC-domain with low order parameters and a DPPC-CHOL-domain with higher order parameters. We have seen that the impact of CHOL on DLiPC domains is very small (see figure \ref{fig:averageS}). As a consequence there is hardly any gain in free energy due to the mutual interaction. In contrast, for DPPC-CHOL domains the DPPC order parameter is strongly increased which implies a significant free energy gain as compared to the pure DPPC domain. Thus, these observations suggests that the free energy is lower if CHOL forms the liquid ordered phase together with DPPC, which is exactly what happens upon raft formation.

\begin{figure}[htbp]
    \centering
    \includegraphics[width=0.9\textwidth]{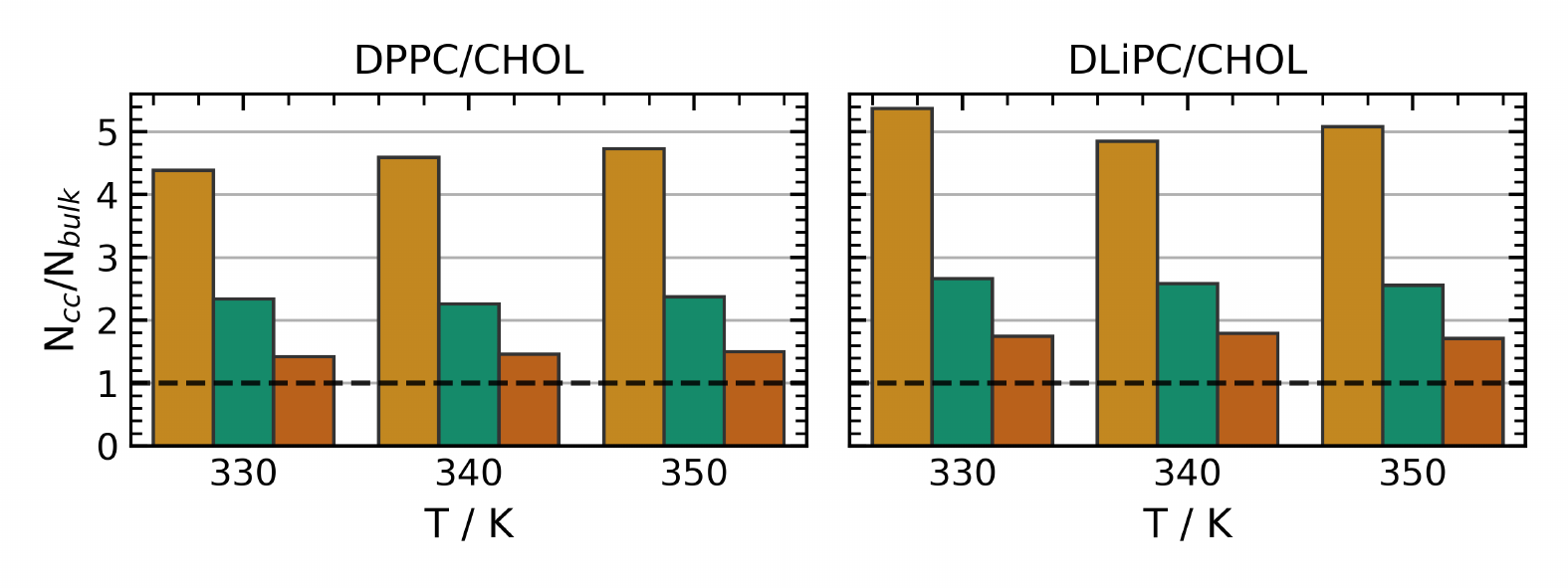}
    \caption{Number of CHOL neighbors of CHOL divided by the number of CHOL neighbors that would arise if CHOL is distributed homogeneously such that values larger than unity indicate CHOL agglomeration. The data is shown for concentrations of 10, 20, and 30\% (yellow, green, and red) in mixtures with DPPC (left) and DLiPC (right).}
    \label{fig:nr_cholchol}
\end{figure}

As another observable we characterize the agglomeration of CHOL. It is reflected by the average number of CHOL neighbors around CHOL which we extract from the CHOL-CHOL RDF for distances smaller than 1 nm. When starting from the RDF, this value is automatically normalized by the number of nearest CHOL neighbors in case of a statistical distribution of CHOL molecules. The results are shown in figure \ref{fig:nr_cholchol}. For all temperatures and concentrations, an  agglomeration of CHOL is observed. For concentrations of 10\%, the enrichment is even 4 to 5 times larger than for the case of a homogeneous CHOL distribution. The CHOL-CHOL enrichment is slightly higher in mixtures with DLiPC, and, with increasing concentration, the agglomeration effect becomes smaller. This may be explained via a simplified view of the bilayers. Envisioning the bilayer as a lattice with a fixed number of sites for PLs and CHOL (e.g., the formulation in figure \ref{fig:squarelattice}), the space for CHOL gets filled upon increasing CHOL concentrations, such that the local concentration becomes similar to the bulk concentration and becomes unity when all CHOL spaces are filled. Within the lattice model this would be the case for 50\% CHOL concentration.  

Naturally, the significant interaction between adjacent CHOL molecules is likely to drive this agglomeration effect which in the considered range of temperatures is basically T-independent. Since this interaction is stronger for the DLiPC-CHOL systems, it naturally emerges that the agglomeration effect is somewhat larger for that case. Closer inspection of the PL-CHOL interaction shows that for fixed overall concentration it (approximately) shows a linear dependence on the number $N_c$ of CHOL neighbors. Thus, the overall contribution of the PL-CHOL interaction to the total interaction energy does not depend on the arrangement of CHOL molecules.

\section{Conclusion}

In this work, we quantified the role of cholesterol in bilayer mixtures with DPPC and DLiPC by studying the neighbor and interaction functions as extracted from MD simulations. It was essential to express these functions in terms of the acyl chain order parameter. They form the core ingredients of a lattice model of PL-CHOL mixtures, reflecting the level I contributions and, thus, in particular the enthalpic part.  Furthermore, from the properties of the neighbor functions we came up with a straightforward suggestion for the topology of the lattice model. In this way a minimalistic visualization of the structure of PL-CHOL mixtures is possible. 

It could be shown that the pair interaction of two phospholipids, which in general depends on both order parameters independently, can be expressed as a function of an effective order parameter which was chosen as the minimum \Smin of both order parameters. Among others the properties of the interaction functions clearly rationalize clustering of CHOL for small CHOL concentrations, as seen in the MD simulations.

Surprisingly, the PL-PL and PL-CHOL interaction functions would suggest that, first, the tendency of the lipid chains to attain higher order parameters is stronger for DLiPC and, second, the impact of CHOL should be stronger for DLiPC as well. This contradicts the actual behavior, seen from the MD simulations. The present analysis now clearly enables the relatation of the observations to the entropic contribution of the resulting free energy, reflecting the tendency of the acyl chain to acquire disordered configurations.

A priori it was not evident that the information about the neighbor and interaction functions, extracted from the MD simulations, can be formulated, within a very good approximation (typically less than  2 kJ/mol),  in such general terms. (1) The neighbor as well as the interaction functions hardly depend on the overall CHOL concentration but rather only on the number of CHOL neighbors. This allows one to use the same functions for a large range of CHOL concentrations. Here we also explored that the residual dependence on the CHOL concentration is to a large extent independent of the order parameter and thus does not impact the thermodynamic properties of the PL-CHOL mixture. (2) The reasonable mapping of the 2D PL-PL interaction function on a 1D function enables a direct interpretation of interaction-driven features such as the driving force of the addition of CHOL on the increase of the acyl chain order parameter. (3) When going from the binary to the ternary mixture, the interaction functions basically remain the same. Thus, there is the realistic hope to apply this approach also to properties such as a closer thermodynamic understanding of raft formation where in different regions of the sample different DPPC/DLiPC/CHOL ratios are found. 

This analysis may naturally serve as the basis for the actual analysis of the lattice model. However, this requires the additional determination of the acyl chain entropy via an iterative procedure in analogy to previous work on cholesterol-free mixtures which requires techniques such as used in previous work on CHOL-free lipid membranes and which, thus, brings in additional degrees of complexity. In this way, a more detailed thermodynamic perspective on the nature of raft formation in these model membranes would be conceivable, since the individual enthalpic and entropic contributions would be explicitly known.

\section{Acknowledgment}
We gratefully acknowledge the SFB 1348 and SFB 858 for funding.

\typeout{}
\bibliography{f_kell07_preprint_acs}

\newpage
\renewcommand{\thesection}{S\arabic{section}}
\renewcommand{\thetable}{S\arabic{table}}
\renewcommand{\thefigure}{S\arabic{figure}}
\setcounter{figure}{0}
\setcounter{table}{0}
\setcounter{page}{1}
%\setcounter{section}{0}

%%%%%%%%%%%%%%%%%%%%%%%%%%%%%%%%%%%%%%%%%%%%%%%%%%%%%%%%%%%%%%%%%%%%%
%% The same is true for Supporting Information, which should use the
%% suppinfo environment.
%%%%%%%%%%%%%%%%%%%%%%%%%%%%%%%%%%%%%%%%%%%%%%%%%%%%%%%%%%%%%%%%%%%%%
\begin{suppinfo}

	%A listing of the contents of each file supplied as Supporting Information
	%should be included. For instructions on what should be included in the
	%Supporting Information as well as how to prepare this material for
	%publications, refer to the journal's Instructions for Authors.

The following files are available free of charge. \\

\section{Computational details}

\begin{table*}[htbp]
\caption{Simulation lengths, temperatures and compositions of the utilized MD simulations.}
\begin{tabular}{ccccc}
System & Temperatures (K) & Lengths ($\mu$s) & \#PL & \#CHOL \\ 
\hline
DPPC & 290-350 & 1 & 350 & 0 \\ 
DLiPC & 290-350 & 1 & 316 & 0 \\ 
DPPC/DLiPC & 290-350 & 1 & 166/166 & 0 \\ 
DPPC/CHOL10\% & 290-350 & 1 & 342 & 38 \\ 
DPPC/CHOL20\% & 290-350 & 1 & 304 & 76 \\ 
DPPC/CHOL30\% & 290-350 & 1 & 280 & 120 \\ 
DLiPC/CHOL10\% & 290-350 & 1 & 306 & 34 \\ 
DLiPC/CHOL20\% & 290-350 & 1 & 280 & 70 \\ 
DLiPC/CHOL30\% & 290-350 & 1 & 266 & 114 \\ 
DPPC/DLiPC/CHOL10\% & 290-350 & 1 & 162/162 & 36 \\ 
DPPC/DLiPC/CHOL20\% & 290-350 & 1 & 148/148 & 74 \\ 
DPPC/DLiPC/CHOL30\% & 290-350 & 1 & 140/140 & 120 \\ 
\end{tabular}
\label{tab:MDlist}
\end{table*}

\begin{table*}[htbp]
\caption{Ensembles and force constants used during the equilibration steps on head group positions and chain dihedrals.}
\begin{tabular}{ccccc}
step & time / ps & ensemble & P atom z pos. (kJ/mol) & chain dih. ang. (kJ/mol) \\ 
\hline
1 & 25 & NVT & 1000 & 1000 \\ 
2 & 25 & NVT & 1000 & 400 \\ 
3 & 25 & NVT & 400 & 200 \\ 
4 & 100 & NpT & 200 & 200 \\ 
5 & 100 & NpT & 40 & 100 \\ 
6 & 100 & NpT & 0 & 0 \\ 
\end{tabular}
\label{tab:equilibration_protocol}
\end{table*}

\begin{figure}[htbp]
    \centering
    \includegraphics[width=0.9\textwidth]{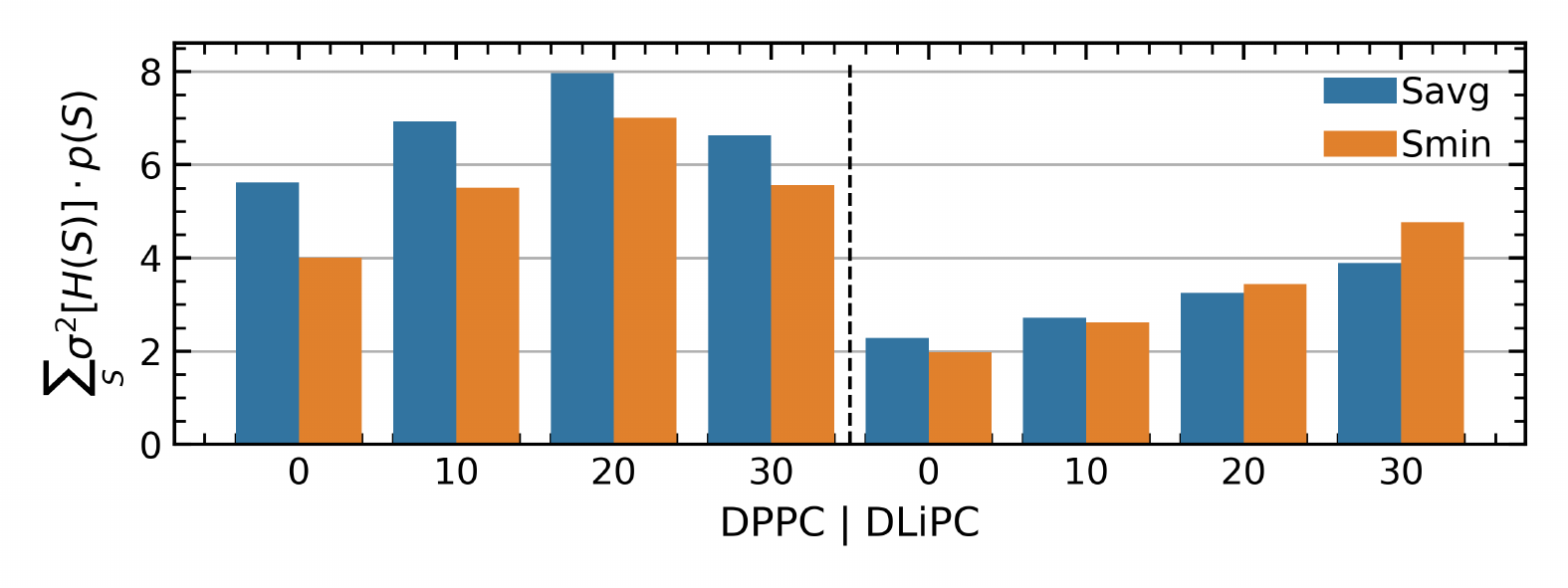}
    \caption{Sum of variances per order parameter bin (see figure \ref{fig:s_scheme_var}) weighted by the respective order parameter distributions for the \Smin and \Savg scheme.}
    \label{fig:s_scheme_var_bar}
\end{figure}

\begin{figure}[htbp]
    \centering
    \includegraphics[width=0.9\textwidth]{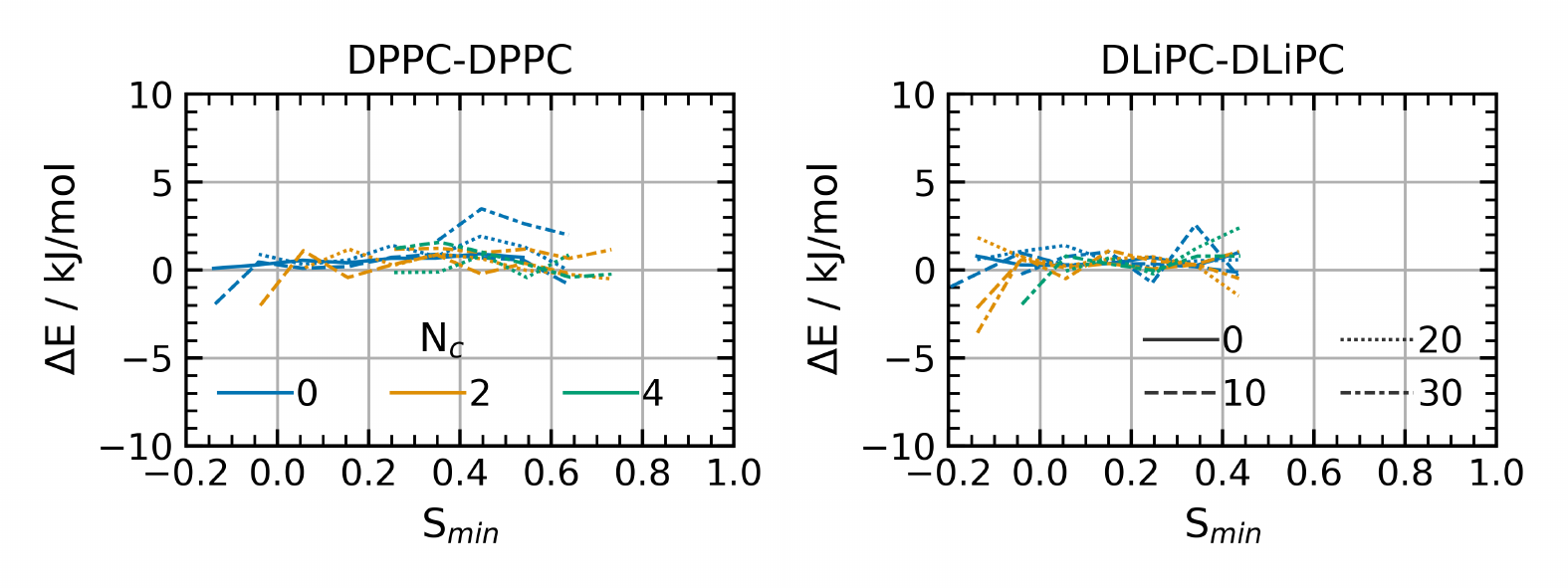}
    \caption{Pair interaction difference between energies from simulations of 330~K and 350~K at CHOL concentrations of 0, 10, 20, and 30~mol\%.}
    \label{fig:deltaEplpl_Tdepscheme}
\end{figure}

\end{suppinfo}

\end{document}